\documentclass[letterpaper,aps,pra,showpacs,twocolumn]{revtex4-1}
\usepackage{amsmath}
\usepackage{graphicx}
\usepackage{float}
\usepackage[ansinew]{inputenc}
\usepackage{array}
\usepackage{color}
\usepackage{amsxtra}
\usepackage{amstext}
\usepackage{amssymb}
\usepackage{latexsym}
\usepackage{gensymb}
\usepackage{dsfont}
\usepackage{braket}
\begin{document}

\title{Feedback-induced Bistability of an Optically Levitated Nanoparticle:\hspace{5cm} A Fokker-Planck Treatment}
\author{Wenchao Ge}
\author{Brandon Rodenburg}
\author{M. Bhattacharya}
\affiliation{School of Physics and Astronomy, Rochester Institute of Technology, 84 Lomb Memorial Drive,
Rochester, NY 14623, USA}

\date{\today}
\begin{abstract}
Optically levitated nanoparticles have recently emerged as versatile platforms for investigating macroscopic
quantum mechanics and enabling ultrasensitive metrology. In this article we theoretically consider two damping
regimes of an optically levitated nanoparticle cooled by cavityless parametric feedback. Our treatment is based
on a generalized Fokker-Planck equation derived from the quantum master equation presented recently and shown
to agree very well with experiment \cite{Rodenburg:16}. For low damping, we find that the resulting Wigner
function yields the single-peaked oscillator position distribution and recovers the appropriate
energy distribution derived earlier using a classical theory and verified experimentally \cite{Gieseler:14nn}.
For high damping, in contrast, we predict a double-peaked position distribution, which we trace to an underlying
bistability induced by feedback. Unlike in cavity-based optomechanics, stochastic processes play a major role in 
determining the bistable behavior. To support our conclusions, we present analytical expressions as well
as numerical simulations using the truncated Wigner function approach. Our work opens up the prospect of 
developing bistability-based devices, characterization of phase-space dynamics, and investigation of the 
quantum-classical transition using levitated nanoparticles.
\end{abstract}

\maketitle
\section{Introduction}
Optical levitation of small particles was first demonstrated by Ashkin in seminal work performed more than
four decades ago \cite{Ashkin:70}. Since then, optical trapping has become a powerful tool for manipulating atoms
\cite{Chu:86}, dielectric particles \cite{Ashkin:86}, and biological systems \cite{Ashkin:87}.  Recently, optically
levitated dielectric microscopic \cite{Li:11,Yin:13, Arita:13} and nanoscopic
\cite{Gieseler:15,MillenJ.:14, Gieseler:14, Gieseler:14nn, Neukirch:15} particles have been cooled
in both cavity-based \cite{Chang:10,Kiesel:13} and cavity-free \cite{Gieseler:12} setups, paving the way for 
macroscopic quantum mechanics \cite{Romero:11, Meystre:13, Chen:13,  Aspelmeyer:14, Tian:16} and
ultrasensitive metrology \cite{Geraci:10,Arvanitaki:13,Moore:14, Ranjit:15}. Parallel to ongoing experiments aimed
at preparing the nanoparticle ground state in a cavityless configuration, a quantum model of parametric feedback
cooling of an optically levitated nanoparticle was presented recently \cite{Rodenburg:16}. This model provided a
master equation, whose predictions regarding feedback cooling agreed very well with experimental data in the
classical regime.

In this article, we derive a Fokker-Planck (FP) equation from the above-mentioned master equation. As is well known,
the FP equation is well-suited for analysing the phase-space distribution for a quantum system, is convenient for
studying the quantum-classical boundary, and is also an efficient calculational tool for dealing with high excitation
numbers which make simulation of the master equation computationally difficult \cite{Carmichael}. In the present
work, we use the FP approach to analyse one of the features which strongly distinguishes parametric feedback cooling
from other active feedback schemes for optomechanical cooling \cite{Mancini:98,Wilson:15,Krause:15}, namely the highly
nonlinear nature of the resultant damping. It was shown theoretically as well as experimentally earlier that such
damping generally results in nonlinear phonon dynamics for - and specifically nonexponential loss of energy from -
the nanoparticle \cite{Rodenburg:16}. Such behavior stands in contrast to most optomechanical systems where the
mechanical element can be described as a linearly damped oscillator even when coupled to optical radiation, and linear
response theory yields a satisfactory description \cite{Aspelmeyer:14}.

We examine two qualitatively different regimes of nanoparticle feedback damping, namely low and high damping, respectively.
In the low damping regime, we obtain a Wigner function which yields a single-peaked position distribution as
well as the energy distribution calculated earlier using a classical theory and verified experimentally
\cite{Gieseler:14nn, Gieseler:15}. In the high damping regime, in contrast, we predict a \emph{double-peaked} position
distribution which we trace to an underlying parametric feedback-induced bistability. Unlike in cavity-based
systems, where bistability can be explained using the mean (drift-like) effects of active
\cite{Mancini:98, Wilson:15, Krause:15, Genes2:08} or passive
\cite{Dorsel:83, Meystre:85, Metzger:08, Mueller:08, Xu:15} feedback, in our case the diffusive processes
present in the system provide a major contribution to bistable behavior \cite{Kus:83}. Our results
open the way to the investigation of levitated cavity-free optomechanical systems in relation to bistability-related
fundamental effects such as squeezing \cite{Genes:08} and entanglement \cite{Ghobadi:11}, and for the construction of
useful devices \cite{Marquardt:06}.

The remainder of this paper is arranged as follows. Section \ref{sec:FPE} introduces the FP equation and recovers
analytically the predictions of the master equation regarding the nanoparticle phonon number. Section \ref{sec:TWD}
introduces the truncated Wigner distribution and reproduces numerically the results of the master equation, showing
the validity of the truncation. Further, it describes analytically and numerically the low and high damping regimes,
respectively, for the levitated nanoparticle. Section \ref{sec:Conc} supplies a conclusion.

\section{The FP equation}
\label{sec:FPE}
\begin{figure}[tbp]
\centering
\includegraphics[width=0.95\columnwidth]{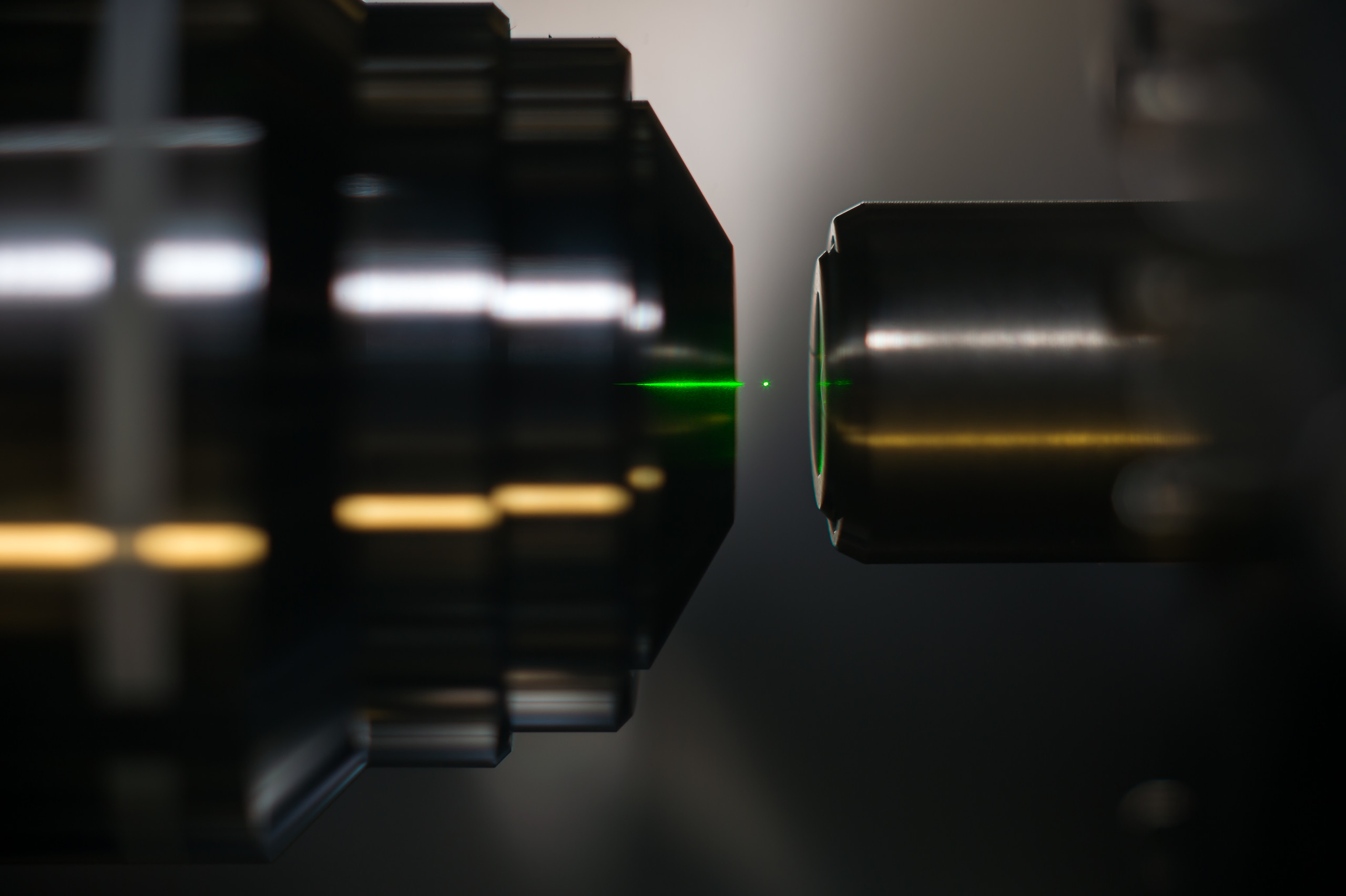}
\caption{Levitated nanoparticle considered in this article. Photograph courtesy of J. Adam Fenster, University of Rochester.
\label{fig:Setup}}
\end{figure}

The system under consideration is a dielectric nanoparticle trapped at the focus of a single Gaussian beam as
described in earlier work \cite{Rodenburg:16, Gieseler:12, Gieseler:15, Neukirch:15}, and shown in Fig.~\ref{fig:Setup}.
The motion of the trapped nanoparticle is detected by an additional probe beam. The detected signal is phase-shifted,
frequency doubled, and then fed back to the trap beam to cool its motion. For small oscillation amplitudes, motion
along each trap direction can be treated separately \cite{Gieseler:14}. The master equation for the nanoparticle
$z$-motion is given by  \cite{Rodenburg:16}
\begin{eqnarray}
\dot{\rho}=&-&i[\omega_z b^{\dagger}_zb_z,\rho]-\left(\frac{A_t+A_p+D_p}{2}\right)\mathcal{D}[Q_z]\rho\nonumber\\
&-&\frac{D_q}{2}\mathcal{D}[P_z]\rho-i\frac{\gamma_g}{2}[Q_z,\{P_z,\rho\}]\nonumber\\
&-&i\gamma_f\left[Q_z^3,\{P_z,\rho\}\right]-\Gamma_f\mathcal{D}[Q_z^3]\rho,
\label{eq:master_com}
\end{eqnarray}
where $\rho$ is the nanoparticle density matrix. The first term on the right hand side of the above equation represents
the harmonic oscillation with trapping frequency $\omega_z$. The second and third terms arise from the diffusion of the
nanoparticle momentum and position, respectively, where $A_t (A_p)$ is the scattering due to the trap (probe) field, and
$D_p=2\gamma_gN_0$ and $D_q=\gamma_g/(6N_0)$ are the coefficients of momentum and the position diffusion, respectively,
due to the background gas. $N_0=k_BT/\hbar \omega_{z} \gg 1$ is the thermal phonon number with $k_B$ the Boltzmann
constant and $T$ the gas temperature. The fourth term describes the damping due to the gas at the rate
$\gamma_g=\eta_{f}/2m$, where $\eta_f$ is the friction force per unit volume due to the gas and $m$ is the mass
of the nanoparticle. The fifth term and the sixth terms correspond to the nonlinear (in oscillator variables) feedback damping
and the corresponding feedback backaction, respectively. The feedback drift and diffusion are characterized by the coefficients
$\gamma_f=\chi^2\Phi G$, and $\Gamma_f=\chi^2 \Phi G^2$, respectively, where $\chi$ is the scaled optomechanical coupling,
$\Phi$ is the average detected flux of the probe photons, and $G$ is the feedback gain. The annihilation (creation)
operator of the nanoparticle motion along $z$ direction is $b_z$ ($b_z^{\dagger}$). The dimensionless position and
momentum operators are $Q_z=b^{\dagger}_z+b_z$ and $P_z=i(b^{\dagger}-b)$, respectively. The Lindblad superoperator in
Eq.~(\ref{eq:master_com}) is $\mathcal{D}[Q_z]\rho=Q_z^{\dagger}Q_z\rho+\rho Q_z^{\dagger}Q_z-2Q_z\rho Q_z^{\dagger}$.

The Wigner function for a quantum system with the density matrix $\rho$ is defined as \cite{SZ}
\begin{eqnarray}
W(X,P)=\frac{1}{\pi \hbar}\int e^{\frac{P\xi}{i\hbar}}\bra{X+\xi/2}\rho\ket{X-\xi/2}d\xi,
\end{eqnarray}
where $\ket{X}$ is the eigenstate of the system at the position $X$. For the dimensionless position operator $Q_z$,
we have $Q_z\ket{X}=\sqrt{2}\frac{X}{x_0}\ket{X}$. For the dimensionless momentum operator $P_z$, we have $P_z\ket{X}=\sqrt{2}\frac{i\hbar}{p_0}\frac{\partial}{\partial X}\ket{X}$. Here $x_0=\sqrt{\frac{\hbar}{m\omega_z}}$
and $p_0=\sqrt{\hbar m \omega_z}$. By applying the position states $\bra{X+\xi/2}$ and $\ket{X-\xi/2}$ from left
and right to Eq.~\eqref{eq:master_com} and performing the integration, we obtain the FP equation for the Wigner
distribution as
\begin{widetext}
\begin{eqnarray}
\frac{\partial W(x,p,t)}{\partial t}&&=\left( -\omega_z p \frac{\partial }{\partial x} +\omega_z x\frac{\partial }{\partial p}
\right)W(x,p,t)+\left[2\gamma_g \frac{\partial }{\partial p
} p+(A_t+A_p+D_p)\frac{\partial ^2}{\partial p^2} + D_q\frac{\partial^2 }{\partial x^2} \right]W(x,p,t)\nonumber\\
&&+\left(24\gamma_f x^2 \frac{\partial }{\partial p} p-2\gamma_f \frac{\partial^3 }{\partial p^3} p\right)W(x,p,t)
+\left(72\Gamma_f x^4 \frac{\partial^2 }{\partial p^2}- 12\Gamma_f x^2\frac{\partial ^4}{\partial p^4} +\frac{\Gamma_f }{2}
\frac{\partial ^6}{\partial p^6}\right) W(x,p,t),
\label{eq:Wigner_d}
\end{eqnarray}
\end{widetext}
where we have defined the dimensionless position and momentum $x=X/x_0$ and $p=P/p_0$, respectively. The first term on
the right-hand-side of Eq.~\eqref{eq:Wigner_d} represents the mechanical oscillation, the second term includes the
gas damping, the gas diffusion and the scattering from optical fields, respectively, the third term originates from the
nonlinear feedback damping, and the fourth term accounts for the diffusion due to feedback backaction. Since
Eq.~\eqref{eq:Wigner_d} is a complicated sixth order partial differential equation, we begin below by first considering
some simple limits before moving on to more complex situations.

\subsection{No feedback}
\label{subsec:NF}
We first consider the case when there is no feedback applied to the optically trapped nanoparticle. In this case, we can set
the feedback gain $G=0$, such that $\gamma_f=\Gamma_f=0$. Written in a compact form, the FP equation for the Wigner function
is then given by
\begin{eqnarray}
\frac{\partial W}{\partial t}=\gamma_{ij}\frac{\partial }{\partial x_i}(x_jW)+D_{ij}\frac{\partial^2W }{\partial x_i\partial x_j},
\label{eq:Wigner_nf}
\end{eqnarray}
where $x_1=x$, $x_2=p$, and the drift matrix $\gamma$ and the diffusion matrix $\bm{D}$ are given by
\begin{eqnarray}
\bm{\gamma}= \left( \begin{array}{cc}
0 & -\omega_z \\
\omega_z & 2\gamma_g
\end{array} \right),
\end{eqnarray}
and
\begin{eqnarray}
\bm{D}= \left( \begin{array}{cc}
D_q & 0\\
0 & A_t+A_p+D_p
\end{array} \right),
\end{eqnarray}
respectively. It can be readily seen by inspection that $D$ is positive definite, i.e.~ that it is symmetric and
its eigenvalues are positive.

For any initial condition, the system reaches a steady-state after a long enough time and then
$\partial W/\partial t=0$. We consider an ansatz of the steady-state solution to the $W$ function as a
Gaussian distribution \cite{HR}, i.e.
\begin{eqnarray}
W_{\text{ss}}(x,p)=\frac{1}{2\pi}\frac{1}{\sqrt{ab-c^2}}e^{-\frac{1}{2(ab-c^2)}(bx^2+ap^2-2cxp)}.
\end{eqnarray}
By substituting the expression of $W_{\text{ss}}(x,p)$ into Eq.~\eqref{eq:Wigner_nf}, we obtain the $a, \ b, \ c$ as
\begin{eqnarray}
a&=&\frac{A_t+A_p+D_p+D_q}{2\gamma_g}+\frac{2\gamma_gD_q}{\omega_z^2},\\
b&=&\frac{A_t+A_p+D_p+D_q}{2\gamma_g},\\
c&=&-\frac{D_q}{\omega_z}.
\end{eqnarray}
From the Wigner function, we obtain the steady-state moments of the nanoparticle as
\begin{eqnarray}
\braket{x^2}_{\text{ss}}&=&\frac{A_t+A_p+D_p+D_q}{2\gamma_g}+\frac{2\gamma_gD_q}{\omega_z^2}, \label{eq:mean_x_nf} \\
\braket{p^2}_{\text{ss}}&=&\frac{A_t+A_p+D_p+D_q}{2\gamma_g},\label{eq:mean_p_nf}\\
\braket{xp}_{\text{ss}}&=&-\frac{D_q}{\omega_z}
\end{eqnarray}
If $(\gamma_{g},D_{q}) \ll \omega_{z}$, which applies experimentally, equipartition is exact and
$\braket{x^2}_{\text{ss}}=\braket{p^2}_{\text{ss}}$. The mean steady-state phonon number of the nanoparticle
without feedback cooling is given by \cite{SZ}
\begin{eqnarray}
\braket{n}_{\text{ss}}&=&\int W_{\text{ss}} \left(\frac{x^2+p^2}{2}\right)dx dp-\frac{1}{2}\nonumber\\
&=&\frac{\braket{x^2}_{\text{ss}}}{2}+\frac{\braket{p^2}_{\text{ss}}}{2}-\frac{1}{2}\nonumber\\
&=&\frac{A_t+A_p+D_p+D_q}{2\gamma_g}+\frac{\gamma_g}{\omega_z^2}D_q-\frac{1}{2},
\label{eq:meanph_nf}
\end{eqnarray}
where $n=b_z^{\dagger}b_z$. This result shows the thermal equilibrium of the particle with its surrounding gas
in the presence of the trapping and probing lasers. We note that in this regime the nanoparticle is linearly
damped by the gas. For high gas pressure and high temperature, Eq.~\eqref{eq:meanph_nf} implies $\braket{n}_{\text{ss}}\approx D_p/(2\gamma_g)=k_BT/(\hbar \omega_z)$.

\subsection{With parametric feedback}
\label{subsec:WPF}
We now consider the case where the feedback is turned on, which cools the nanoparticle into a lower mean phonon
number state. In this case we were unable to solve for the Wigner function analytically. Instead, we employed
the full FP equation in Eq.~\eqref{eq:Wigner_d} to generate phase-space moments of the nanoparticle. We multiplied
system variables with Eq.~\eqref{eq:Wigner_d} and integrated over phase-space to obtain the following relevant
equations
\begin{eqnarray}
\braket{\dot{x}^2}&=&2\omega_z \braket{xp}+2D_q,\label{eq:15}\\
\braket{\dot{p}^2}&=&-2\omega_z \braket{xp}-4\gamma_g\braket{p^2}+2(A_t+A_p+D_p)\nonumber\\
&&+144\Gamma_f\braket{x^4}-48\gamma_f\braket{x^2p^2},\label{eq:16}\\
\braket{\dot{xp}}&=&\omega_z\braket{p^2}-\omega_z\braket{x^2}-2\gamma_g\braket{xp}\nonumber\\
&&-24\gamma_f\braket{x^3p},\label{eq:17}\\
\braket{\dot{x^3p}}&=&3\omega_z\braket{x^2p^2}-\omega_z\braket{x^4}-2\gamma_g\braket{x^3p}\nonumber\\
&&-24\gamma_f\braket{x^5p},\label{eq:x3p}\\
\braket{\dot{x}^4}&=&4\omega_z \braket{x^3p}+12D_q\braket{x^2},\label{eq:19}\\
\braket{\dot{x}^6}&=&6\omega_z \braket{x^5p}+30D_q\braket{x^4},
\label{eq:Moments}
\end{eqnarray}
where for any moment $f(x,p)$, $\braket{f(x,p)}=\int W(x,p,t)f(x,p)dxdp$ and we have used the boundary conditions
$W(\pm \infty,p,t)=0$ and $W(x, \pm \infty,t)=0$. Equations \eqref{eq:15}--\eqref{eq:Moments} are not a closed set of equations
for all involved moments, but as we will see below, they contain all the information required for solving for the
steady state phonon number. In the steady-state, we obtain
\begin{eqnarray}
&&\braket{p^2}_{\text{ss}}=\left(1-72\frac{\gamma_fD_q}{\omega_z^2}\right)\braket{x^2}_{\text{ss}}
-\frac{2\gamma_gD_q}{\omega_z^2},\label{eq:mean_p}\\
&&A\braket{x^4}_{\text{ss}}+B\braket{x^2}_{\text{ss}}+C=0\label{eq:mean_x},
\end{eqnarray}
with the coefficients $A=8\gamma_f\left( 1-9\frac{\Gamma_f}{\gamma_f}-120\frac{\gamma_fD_q}{\omega_z^2}\right)$, $B=2\gamma_g\left(1-96\frac{\gamma_fD_q}{\omega_z^2}\right)$, and $C=A_t+A_p+D_p+D_q+\frac{4\gamma_g^2D_q}{\omega_z^2}$.
Comparing Eq.~\eqref{eq:mean_p} with Eqs.~\eqref{eq:mean_x_nf} and \eqref{eq:mean_p_nf} we see that there is an additional
deviation from exact equipartition due to the parametric feedback. For typical experimental parameters, $\gamma_{f} \ll
\omega_{z}$ , and this deviation is quite small.



In order to solve Eq.~\eqref{eq:mean_x}, we further assume that the position is described approximately by a zero-mean
Gaussian distribution (as for a thermal state), which gives the relation $\braket{x^4}=3\braket{x^2}^2$. After some
rearrangement Eq.~\eqref{eq:mean_x} can be rewritten as
\begin{eqnarray}
2J\braket{x^2}_{\text{ss}}^2+2\gamma_g\braket{x^2}_{\text{ss}}-(A_t+A_p+D_p)=0,
\label{eq:mean_x2}
\end{eqnarray}
where $J=12(\gamma_f-9\Gamma_f)$ and we have neglected all the terms proportional to $D_q$ since $D_q\ll \omega_z$
in the experiment. We obtain from Eq.~\eqref{eq:mean_x2} the steady-state position-squared mean as
\begin{eqnarray}
\braket{x^2}_{\text{ss}}&=&-\frac{\gamma_g}{2J}+\frac{\sqrt{\gamma_g^2+2J(A_t+A_p+D_p)}}{2J}\nonumber\\
&\approx&\sqrt{\frac{A_t+A_p+D_p}{2J}},
\label{eq:x2_wigner}
\end{eqnarray}
where the approximation is valid for $J N_0\gg\gamma_g$. Therefore, using the equipartition condition obtained from
Eq.~\eqref{eq:mean_p}, we obtain
\begin{eqnarray}
\braket{n}_{\text{ss}}&=&\braket{x^2}_{\text{ss}}-\frac{1}{2}\nonumber\\
&\approx&\sqrt{\frac{A_t+A_p+D_p}{2J}}-\frac{1}{2},\label{eq:steady_n}
\end{eqnarray}
This result agrees with that in Eq. (14b) of Ref.~\cite{Rodenburg:16} obtained using the master equation
Eq.~\eqref{eq:master_com} [the $-1/2$ term was neglected in going from Eq.(14a) to Eq.(14b)]. For higher-order
moments, such as $\braket{x^4}_{\text{ss}}$, required for calculating correlation functions such as $g^{(2)}$,
it is difficult to solve the equations of the moments since they involve even higher-order moments. We will
calculate those quantities using numerical and analytical solutions to the FP equation in the next section.

\section{Truncated Wigner distribution}
\label{sec:TWD}
In order to study the partial differential equation Eq.~\eqref{eq:Wigner_d} in the presence of feedback in more
detail, we apply an approximation to the FP equation. This approximation consists of the method of truncated Wigner
distributions that has been studied previously \cite{Werner:97}. It allows us to keep up to the second-order
derivatives in the FP equation, thus retaining both the essential deterministic and stochastic parts.

The justification of the truncation can be understood in two limiting cases. In the first case, when the steady-state
phonon number $\braket{n}_{\text{ss}}\gg1$, we apply the method of system size expansion to truncate higher-order
derivatives \cite{Carmichael}. The central idea is that as the system size (measured by the variable
$\braket{n}_{\text{ss}}$ in the present case) increases, the higher-order derivatives can be expanded in terms of a
small parameter related to the inverse of the system size. In that case the system variables, 
$x$, $p\propto \sqrt{\braket{n}_{\text{ss}}}\gg1$, and the higher-order derivatives such as $\partial^4/\partial x^4$
provide much smaller contributions compared to the lower-order derivatives on average.

In the second case, when the steady-state phonon number $\braket{n}_{\text{ss}}<1$, we consider the uncertainty
principle, i.e.~ $\Delta x \Delta p\ge 1/2$ to make the truncation. Using this principle for $\braket{n}_{\text{ss}}<1$,
higher-order derivatives such as $ x^2\partial^4/\partial p^4$ and $\partial ^6/\partial p^6$ can be
shown to be of the same order as $x^4 \partial^2 /\partial p^2$. In Eq.~\eqref{eq:Wigner_d}, we therefore
retain only the contribution with the coefficient $72\Gamma_{f}$ from the last term. The truncated FP equation is
then given by
\begin{eqnarray}
&&\frac{\partial W(x,p,t)}{\partial t} \nonumber\\
&&=\left( -\omega_z p \frac{\partial }{\partial x} +\omega_z x\frac{\partial }{\partial p}
\right)W(x,p,t)\nonumber\\
&&+\left[2\gamma_g \frac{\partial }{\partial p
} p+(A_t+A_p+D_p)\frac{\partial ^2}{\partial p^2} + D_q\frac{\partial^2 }{\partial x^2} \right]W(x,p,t) \nonumber\\
&&+\left(24\gamma_f x^2 \frac{\partial }{\partial p} p+72\Gamma_f x^4 \frac{\partial^2 }{\partial p^2}\right)W(x,p,t).
\label{eq:Wigner_tr}
\end{eqnarray}
Below, we will first recover the results of the master equation analysis of Ref.~\cite{Rodenburg:16} to show that
the truncated FP approach is valid in the range of phonon numbers used in the present article. We will subsequently
present new analytical as well as numerical results based on this equation in the low and high damping regimes.

\subsection{Feedback cooling}
\begin{figure}[tbp]
\centering
\includegraphics[width=0.9\columnwidth]{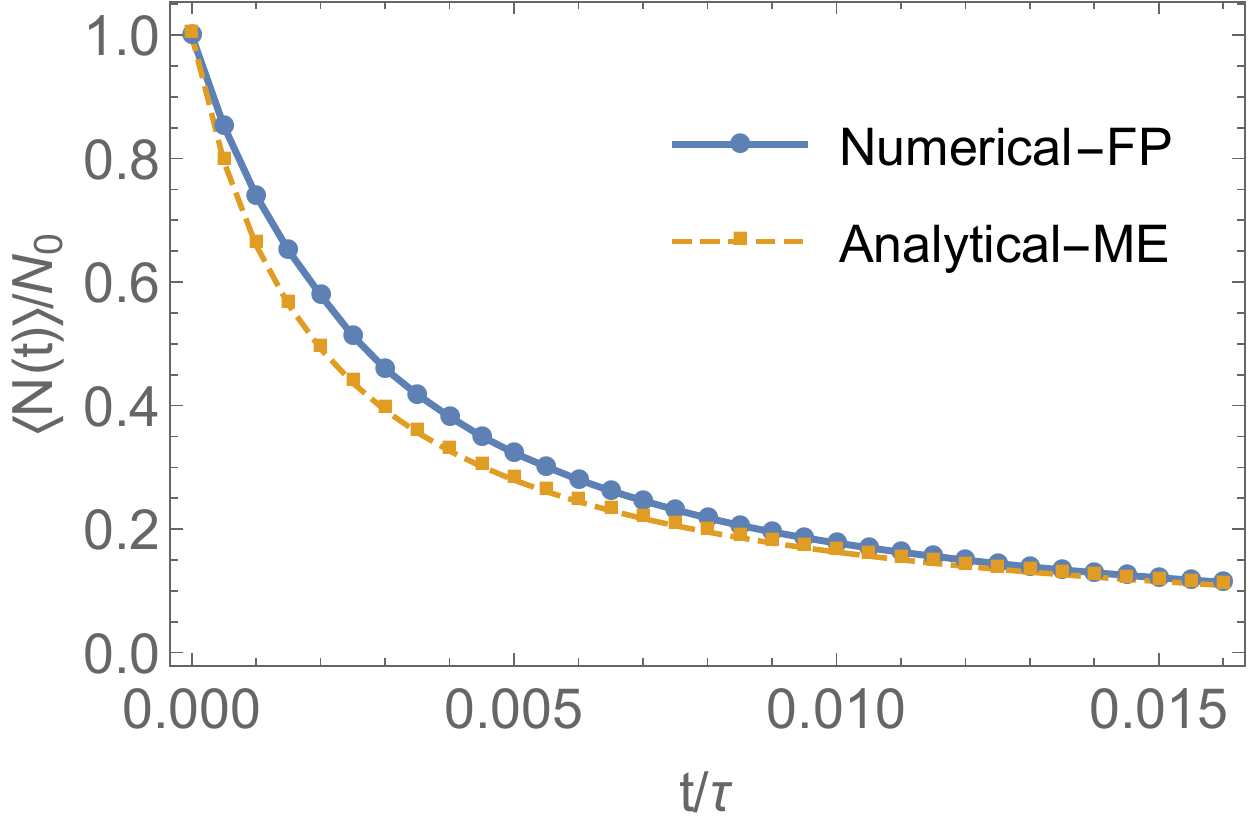}
\caption{The scaled phonon number using the numerical simulation of Eq.~\eqref{eq:Wigner_tr} (solid) and the analytical 
result in Ref. \cite{Rodenburg:16} (dashed) versus time during cooling for initial parameters $\omega_z=40$ kHz,  
$\gamma_g=5$ Hz, $A_t+A_p=1$ kHz, $\gamma_f=18\Gamma_f=2/9$ kHz, and $N_0=10^6$. $\tau$ is the characteristic time of 
the nonlinear feedback cooling given in the text. \label{fig:d_cool}}
\end{figure}

In this section we show that our truncated FP approach agrees with earlier results deduced from the master equation
\cite{Rodenburg:16}. First, we study the dynamical feedback cooling with the FP equation Eq.~\eqref{eq:Wigner_tr}.
In Fig.~\ref{fig:d_cool},
we plot the dynamical phonon number versus $t/\tau$ using both the numerical simulation from the FP equation and the
analytical result in Ref.~\cite{Rodenburg:16} obtained with the master equation. Here the characteristic cooling time $\tau=2\sqrt{(2J+2\gamma_g)^2+8J(A_t+A_p+D_p-J/2)}$.  We observe quite good agreement between the two methods.

\begin{figure}[tbp]
\centering
\includegraphics[width=0.95\columnwidth]{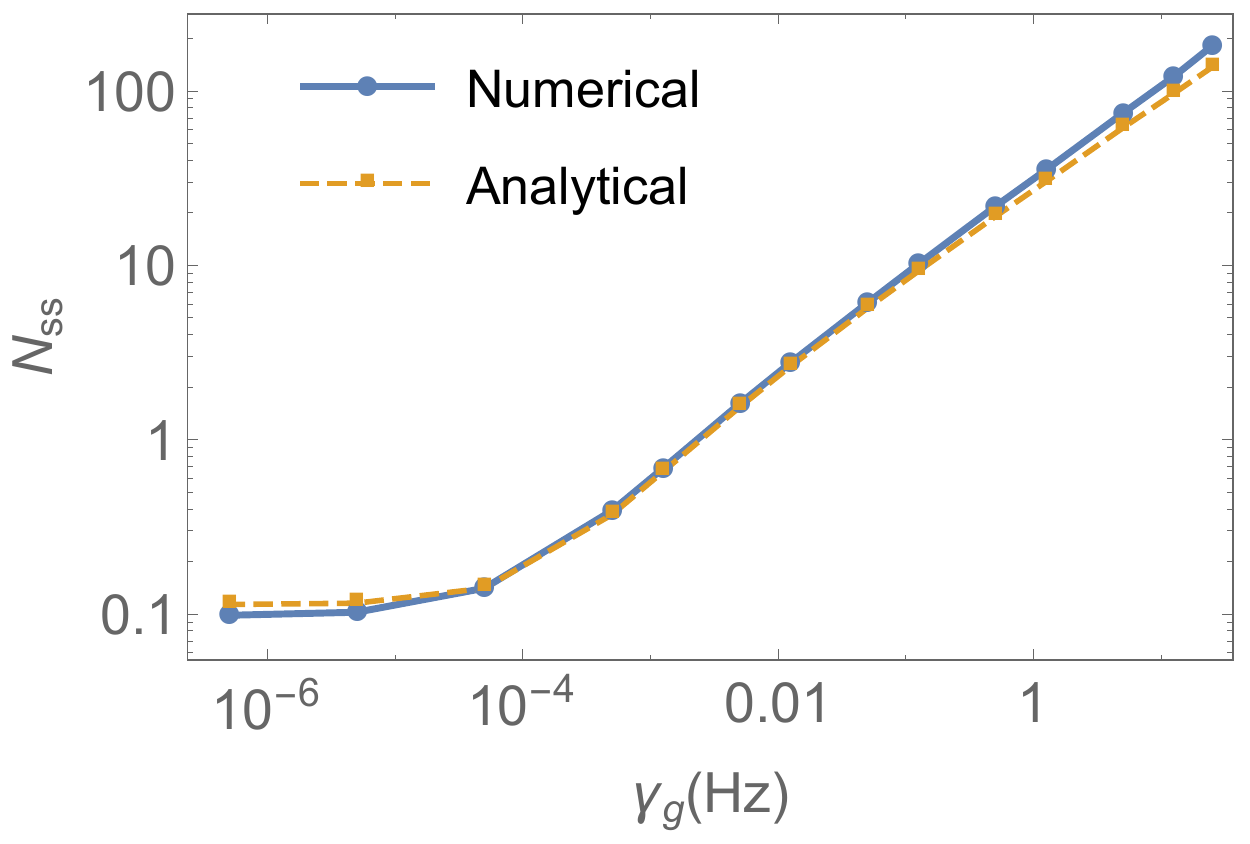}
\caption{Steady-state feedback cooling vs background gas damping rate for using the numerical simulation from 
Eq.~\eqref{eq:Wigner_tr} (solid) and the analytical expression in Eq.~\eqref{eq:steady_n} (dashed). The other parameters 
are the same as in Fig.~\ref{fig:d_cool}.
\label{fig:cool_2}}
\end{figure}

Next, we study the steady-state phonon number for an initial phonon number $N_0=10^6$ for different values
of the background gas scattering rate. As shown in Fig.~\ref{fig:cool_2},
we find good agreement between our numerical results and the earlier analytical result $\braket{n}_{\text{ss}}$.
We observe that steady-state phonon number decreases as $\gamma_g$ is lowered and saturates (i.e.~ becomes
independent of $\gamma_{g}$) for very small $\gamma_g$.

The agreement between calculations using the numerical truncated Wigner distribution method [Eq~\eqref{eq:Wigner_tr}]
and the master equation [Eq.~\eqref{eq:master_com}] occurs over the entire range of phonon numbers considered in this
article and shows the validity of the truncation approximation used by us.

\subsection{Low-damping limit}
\begin{figure}[tbp]
\centering
\includegraphics[width=.9\linewidth]{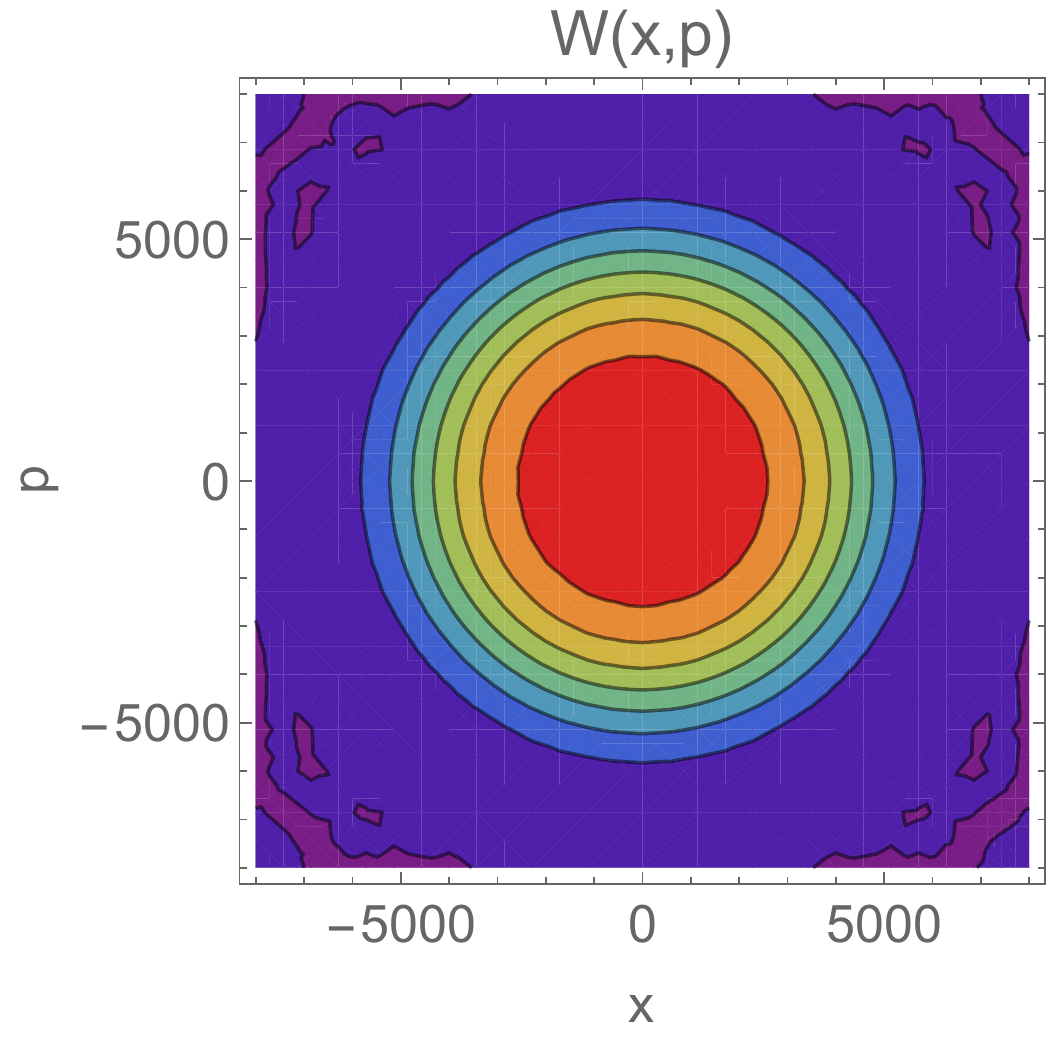}
\caption{Steady-state Wigner distribution calculated numerically in the low damping regime for the $\gamma_g=0.5$ mHz,
$A_t+A_p=1$ kHz, $\gamma_{\text{eff}}=4.45\times10^{-7}$ and $N_0=10^8$. The steady-state phonon number is $\braket{n}_{\text{ns}}=6.68\times10^6$.}
\label{fig:Wigner_6e6}
\end{figure}

We now consider the case when both the gas damping rate and the parametric feedback damping rate are
much smaller than the mechanical oscillation frequency. From Eq.~\eqref{eq:Wigner_tr}, it can be seen
that this limit corresponds to the condition
\begin{eqnarray}
\frac{2\gamma_g+24\gamma_f\braket{x^2}_{\text{ss}}}{\omega_z}\approx \gamma_{\text{eff}}\ll 1,
\label{eq:LowD}
\end{eqnarray}
where we have considered the low-gas-pressure limit in the second step, and
\begin{equation}
 \gamma_{\text{eff}}=\frac{24\gamma_f\braket{x^2}_{\text{ss}}}{\omega_z}.
\end{equation}
Solving Eq.~\eqref{eq:Wigner_tr} numerically in this regime, we obtain the Wigner function shown in 
Fig.~\ref{fig:Wigner_6e6}.
We observe only a single peak in the Wigner distribution well localized near the origin due to the parametric
feedback.

To analyze the phase-space distribution further, we define the dimensionless energy of the system as
$\epsilon=(x^2+p^2)/2$, where $x=\sqrt{2\epsilon}\cos(\omega_z t)$ and
$p=\sqrt{2\epsilon}\sin(\omega_z t)$. Since we have assumed the harmonic oscillation frequency of the nanoparticle
to be much higher than the damping rate, averaging on the timescale of $1/\omega_z$, yields
$x^2\approx p^2\approx \epsilon$. For $\braket{x^2}_{\text{ss}}\gg1$, the low feedback damping implied by
Eq.~\eqref{eq:LowD} corresponds to small feedback gain, i. e., $G\ll 1$, and we can neglect the feedback backaction
term in Eq.~\eqref{eq:Wigner_tr}. With these conditions, we can write the truncated FP equation in phase-space in
terms of the energy variable as
\begin{eqnarray}
&&\frac{\partial W_{\epsilon}(\epsilon,t)}{\partial t} \nonumber\\
&&\approx\frac{\partial }{\partial \epsilon}\left[2\gamma_g \left(\epsilon
-\frac{A_t+A_p+D_p}{2\gamma_g}\right)+12\gamma_f \epsilon^2 \right]W_{\epsilon}(\epsilon,t) \nonumber\\
&&+ (A_t+A_p+D_p)\frac{\partial^2 }{\partial \epsilon^2}\epsilon W_{\epsilon}(\epsilon,t),
\label{eq:Wigner_en}
\end{eqnarray}
where we have dropped both the feedback backaction and the position diffusion terms, and also assume
$x^2p^2\approx \epsilon^2/2$ when averaged over a period of mechanical oscillation. We were able to solve
Eq.~\eqref{eq:Wigner_en} analytically in the steady state for the energy distribution
\begin{eqnarray}
W_{\epsilon \text{ss}}(\epsilon)=\mathcal{N}\exp\left[-\frac{2\gamma_g}{A_t+A_p+D_p}\left(\epsilon+\frac{3\gamma_f }{\gamma_g}\epsilon^2\right)\right],
\label{eq:Wigner_low}
\end{eqnarray}
where $\mathcal{N}$ is a normalization constant. The FP equation [Eq.~\eqref{eq:Wigner_en}] and its steady state
solution [Eq.~\eqref{eq:Wigner_low}] for the energy are exactly the same as obtained earlier using a classical
analysis and verified experimentally \cite{Gieseler:14nn, Gieseler:15}. The origin of the deviation from a
Boltzmann distribution can be seen clearly to be the nonlinear parametric feedback cooling in Eq.~\eqref{eq:Wigner_low}.
The steady state phonon number can be calculated analytically from Eq.~\eqref{eq:Wigner_low} as
$\braket{n}=\int \epsilon W_{\epsilon \text{ss}}(\epsilon)d\epsilon$, but the expression is rather cumbersome and we
do not reproduce it here. Calculating the mean phonon number using both the analytical expression derived from Eq.~
\eqref{eq:Wigner_low} and the numerical simulation for the parameters of Fig.~\ref{fig:Wigner_6e6}, gives the
same result, $\braket{n_{\text{ns}}}=6.68\times10^6$.

\begin{figure}[tbp]
\centering
\includegraphics[width=0.95\columnwidth]{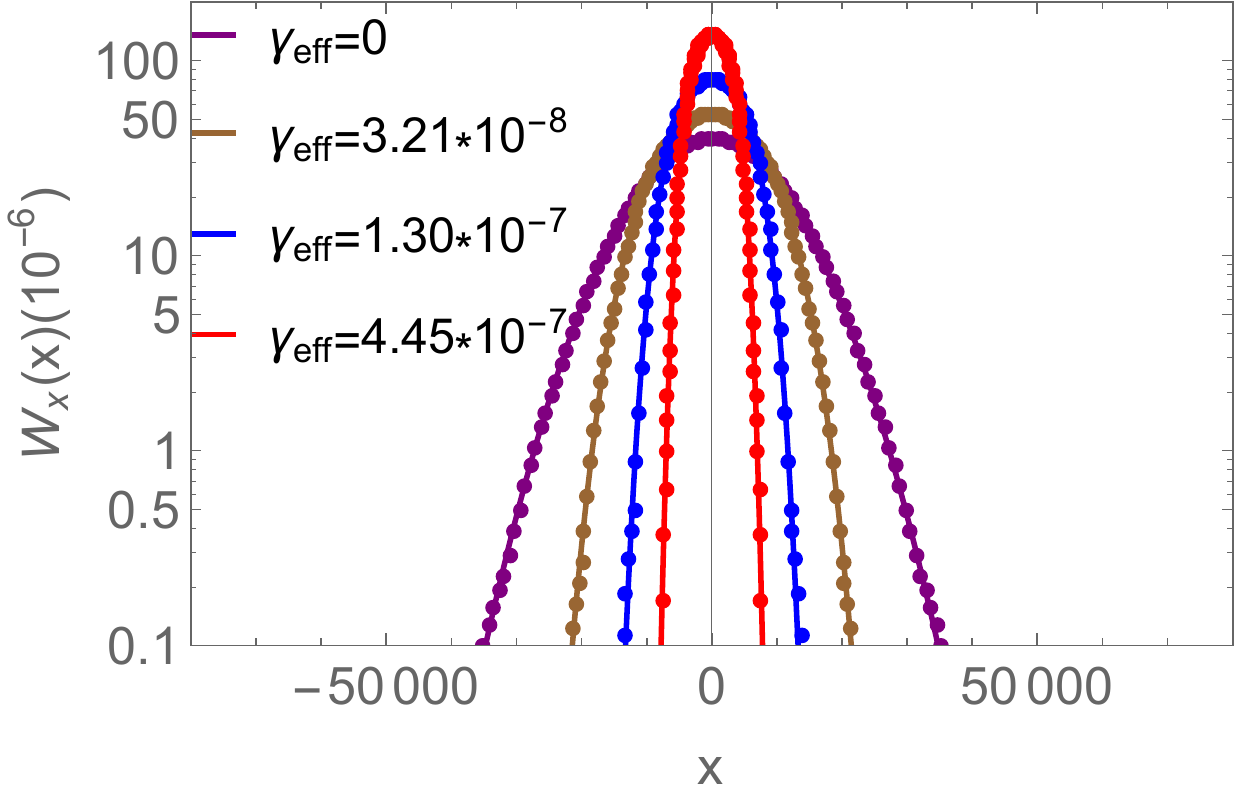}
\caption{Position distributions for different feedback strengths $\gamma_{\text{eff}}$ using
both the analytical expression (solid lines)[Eq.~\eqref{eq:low_x}] and the numerical simulation (dotted symbols).
The remaining parameters are the same as in Fig.~\ref{fig:Wigner_6e6}.
\label{fig:Wx_weak}}
\end{figure}

By replacing $\epsilon=(x^2+p^2)/2$ and performing an integration on the momentum $p$ in Eq.~\eqref{eq:Wigner_low},
we obtain an analytical expression for the position distribution in this low-damping limit,
\begin{eqnarray}
W_{\epsilon \text{ss}}(x)&=&\mathcal{N'}\sqrt{3x^2\gamma_f+\gamma_g}\exp\left[-\frac{(3x^2\gamma_f
+\gamma_g)^2}{12\gamma_f(A_t+A_p+D_p)}\right]\nonumber\\
&&\times K_{\frac{1}{4}}\left[\frac{(3x^2\gamma_f+\gamma_g)^2}{12\gamma_f(A_t+A_p+D_p)}\right],
\label{eq:low_x}
\end{eqnarray}
where  $\mathcal{N'}$ is a normalization constant and $K_{\frac{1}{4}}$ is a generalized Bessel function of
the second kind. This expression for the position distribution agrees with that derived classically in Ref.
\cite{Gieseler:14nn,Gieseler:15}, in the absence of driving, nonlinearity and feedback backaction. We plot the 
position distribution using both the analytical expression of Eq.~\eqref{eq:low_x} and the full numerical simulation of
Eq.~\eqref{eq:Wigner_tr} in Fig.~\ref{fig:Wx_weak}.
As can be seen, both methods agree well with each other, and also with Fig.1 in Ref.~\cite{Gieseler:15}.

To conclude, in this section we demonstrated that in the low damping limit, our quantum calculations show
agreement with earlier classical calculations, which now provides us with a tool for examining the quantum-to-classical
transition. This agreement also demonstrates that our truncation of the FP equation is valid in the low-damping limit.

\subsection{Overdamped limit}
\begin{figure}[tbp]
\centering
\includegraphics[width=0.85\linewidth]{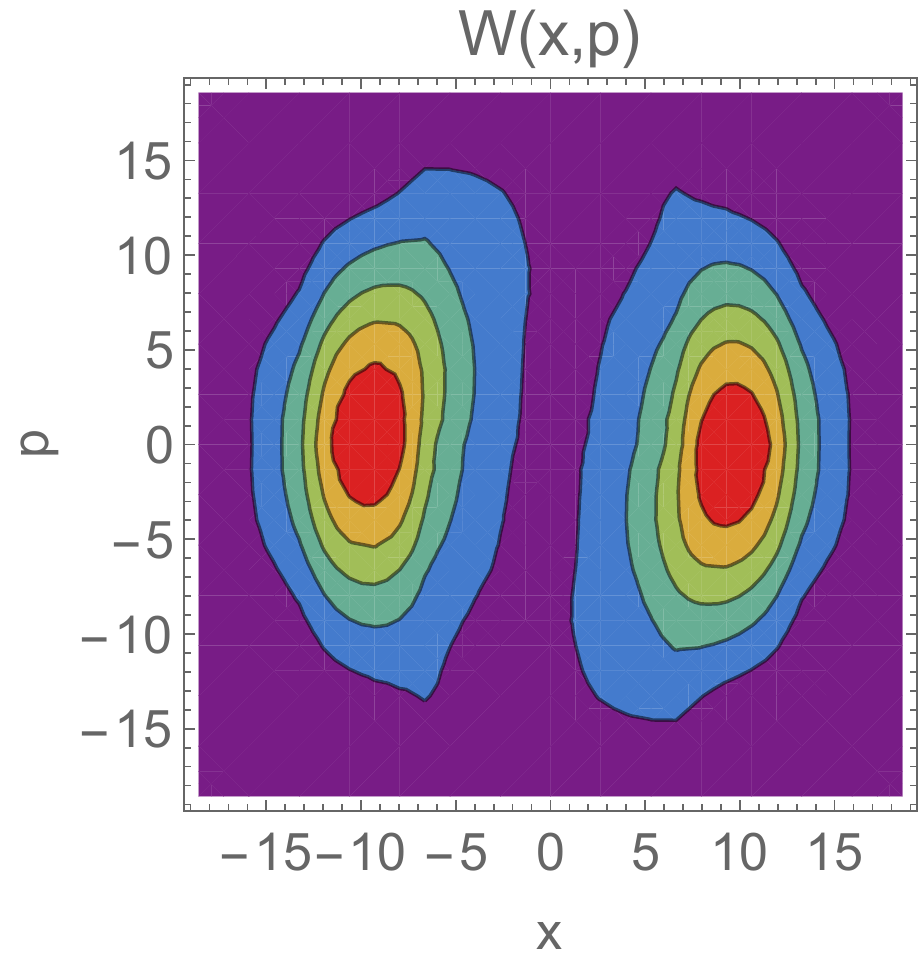}
\caption{Steady-state Wigner distribution obtained numerically in the high feedback damping regime for $\gamma_g=0.05$ Hz, $\gamma_f=27\Gamma_f=4/27$ kHz,
and $N_0=10^8$. The other parameters are the same as in Fig.~\ref{fig:d_cool}. The steady-state phonon number is $\braket{n}_{\text{ns}}\approx84$.}
\label{fig:Wigner_64}
\end{figure}

We now consider the regime in which the nanoparticle motion is overdamped. The numerical solution of
Eq.~\eqref{eq:Wigner_tr} yields a double-peaked Wigner function as shown in Fig.~\ref{fig:Wigner_64}.
This behavior is unlike that in the low-damping regime, as can be seen readily by comparing 
Figs.~\ref{fig:Wigner_6e6} and \ref{fig:Wigner_64}. It is also distinct from earlier bimodal distributions
which were observed in the presence of external driving of the nanoparticle \cite{Gieseler:14nn}; in our case,
there is no such driving.

In order to investigate further, we consider analytical solutions
to Eq.~\eqref{eq:Wigner_tr}. However, even with the truncation of higher-order derivatives in the Wigner
distribution, the FP equation [Eq.~\eqref{eq:Wigner_tr}] is difficult to solve analytically because of the
presence of the nonlinear feedback. We find it is more convenient to first consider the corresponding quantum
Langevin equations for the nanoparticle derived from Eq.~\eqref{eq:Wigner_d} the full FP equation \cite{HR}
\begin{eqnarray}
\dot{x}&=&\omega_z p+F_x(t),\label{eq:x_QL} \\
\dot{p}&=&-\omega_z x-2\gamma_g p-24\gamma_f x^2 p+F_p(t),
\label{eq:p_QL}
\end{eqnarray}
where the correlations of the noises are given explicitly as
\begin{eqnarray}
\braket{F_x(t)F_x(t^{\prime})}&=&2D_q\delta(t-t^{\prime}),\\
\braket{F_p(t)F_p(t^{\prime})}&=&\left[2(A_t+A_p+D_p)+144\Gamma_f x^4\right]\delta(t-t^{\prime}).\nonumber\\
\end{eqnarray}
In the overdamped regime, the condition
\begin{eqnarray}
\frac{2\gamma_g+24\gamma_f\braket{x^2}_{\text{ss}}}{\omega_z}\approx \gamma_{\text{eff}}\gtrsim 1,
\label{eq:overdampc}
\end{eqnarray}
applies. For experimentally accessible parameters \cite{Rodenburg:16}, $\gamma_f\ll\omega_z$ which gives
$\braket{x^2}_{\text{ss}}\gg1$ as the condition for overdamping. We consider the situation
$(2\gamma_g +24\gamma_f x^2) p\gg \dot{p}$ in Eq.~\eqref{eq:p_QL}, which implies $ \dot{p}\approx0$, and
allows us to eliminate the momentum adiabatically. We then obtain from Eq.~\eqref{eq:p_QL}
$p\approx-(\omega_z x-F_p(t))/(2\gamma_g+24\gamma_f x^2)$. Substituting into Eq.~\eqref{eq:x_QL}, the
quantum Langevin equation for the nanoparticle's position can be approximated as
\begin{eqnarray}
\dot{x}&\approx&-\frac{\omega_z^2 x}{2\gamma_g+24\gamma_f x^2}+\frac{\omega_zF_p(t)}{2\gamma_g+24\gamma_f x^2},
\nonumber\\
&=&h(x,t)+g(x,t)\Gamma(t),
\label{eq:sm}
\end{eqnarray}
where we have dropped the position fluctuation force $F_x(t)$, which is negligible because $D_{q} \ll \omega_{z}$
and define
\begin{eqnarray}
h(x,t)&=&-\frac{\omega_z^2 x}{2\gamma_g+24\gamma_f x^2},\\
g(x,t)&=&\frac{\omega_z\sqrt{A_t+A_p+D_p+72\Gamma_f x^4}}{2\gamma_g+24\gamma_f x^2},
\end{eqnarray}
and $\braket{\Gamma(t)\Gamma(t^{\prime})}=2\delta(t-t^{\prime})$.

\subsubsection{The position distribution}
According to Ref. \cite{HR}, the FP equation for the
position of the nanoparticle corresponding to Eq.~\eqref{eq:sm} is given by
\begin{eqnarray}
\frac{\partial W_x(x,t)}{\partial t}=-\frac{\partial }{\partial x}\left(D_x^{(1)}W_x(x,t)\right)+\frac{\partial^2 }{\partial x^2}\left(D_x^{(2)}W_x(x,t)\right),\nonumber\\
\label{eq:smx}
\end{eqnarray}
where $D_x^{(1)}=h(x,t)+\frac{\partial g(x,t)}{\partial x}g(x,t)$, $D_x^{(2)}=\left(g(x,t)\right)^2$, and
$W_x(x,t)=\int W(x,p,t) dp$. In the steady-state, Eq.~\eqref{eq:smx} can be solved analytically to yield the
position distribution
\begin{eqnarray}
\label{eq:Wignerx}
W_{x\text{ss}}(x)&=&\mathcal{N}_{x}\frac{\gamma_g+12x^2\gamma_f}
{\left(A_t+A_p+D_p+72\Gamma_fx^4\right)^{\frac{\gamma_f+6\Gamma_f}{12\Gamma_f}}}\\
&\times&\exp\left[-\frac{\gamma_g\arctan{\left(6\sqrt{2}x^2\sqrt{\frac{\Gamma_f}{A_t+A_p+D_p}}\right)}}
{6\sqrt{2\Gamma_f(A_t+A_p+D_p)}}\right],\nonumber
\end{eqnarray}
where $\mathcal{N}_{x}$ is a normalization constant.

\begin{figure}[tbp]
\centering
\includegraphics[width=0.95\columnwidth]{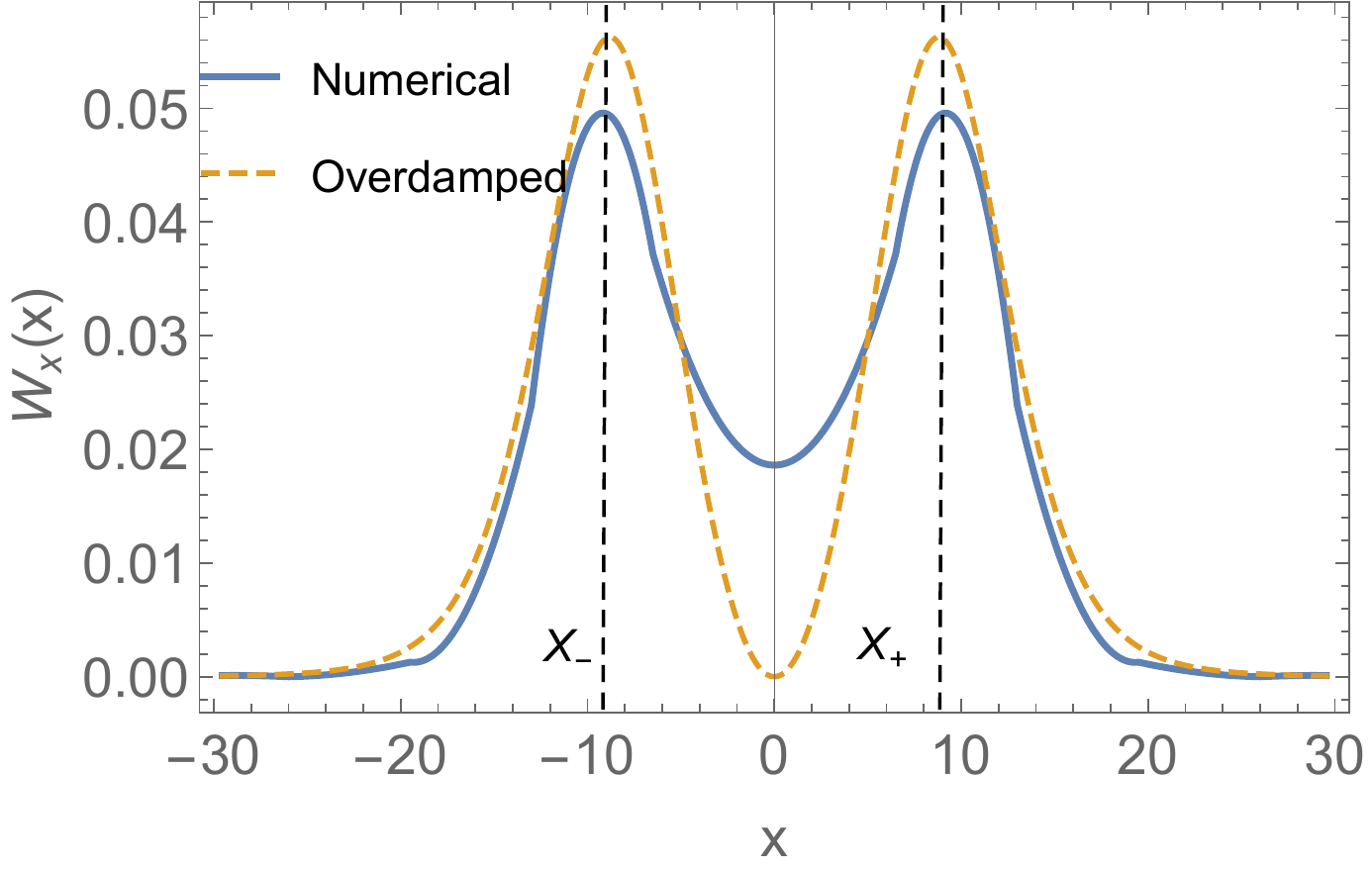}
\caption{Steady-state position distributions from the numerical solution (solid) of Eq.~\eqref{eq:Wigner_tr} and that using
$W_{x\text{ss}}(x)$ [Eq.~\eqref{eq:Wignerx}]  (dashed) for the same parameters in Fig.~\ref{fig:Wigner_64}.
\label{fig:Wx}}
\end{figure}

In Fig.~\ref{fig:Wx}, we have plotted the analytical expression for $W_{x\text{ss}}(x)$ from Eq.~\eqref{eq:Wignerx} as well as
the numerical solution from Eq.~\eqref{eq:Wigner_tr}. Both results show two peaks, which is indicative of bistability. To investigate further, we consider the condition $\gamma_g\ll \sqrt{\Gamma_f(A_t+A_p+D_p)}$ for which the exponential term in Eq.~\eqref{eq:Wignerx} can be dropped as it is almost unity for the range of the position distribution. This condition is equivalent to
$N_0\Gamma_f\gg\gamma_g$ which is valid for the parameter regime we study here. Then the position distribution reduces
to
\begin{equation}
W_{x\text{ss}}(x) \simeq \frac{12 \mathcal{N}_{x}\gamma_{f}x^2}{\left(A_t+A_p+D_p+72\Gamma_fx^4\right)^{\frac{\gamma_f+
6\Gamma_f}{12\Gamma_f}}}.
\label{eq:Wpeaks}
\end{equation}
From this expression we see that the term $\propto \gamma_{f} x^{2}$ in the numerator, which corresponds to feedback
cooling, pushes the particle away from the origin $x=0$, by making the position distribution vanish at that point.
This can be understood intuitively from the fact that the feedback force is $\propto -x^{2}p$ \cite{Neukirch:15},
which vanishes at the origin and allows the particle to transit that point without slowing down. In contrast, the
term $\propto \Gamma_f x^4$ in the denominator of Eq.~\eqref{eq:Wpeaks}, which is due to feedback backaction noise,
pushes the particle towards the origin, by making the position distribution there large. We may therefore expect
equilibria away from the origin. These can be found by taking the derivative of Eq.~\eqref{eq:Wpeaks} with respect
to $x$, and we obtain two maxima corresponding to
\begin{eqnarray}
x_{\pm}=\pm\left(\frac{A_t+A_p+D_p}{2J}\right)^{\frac{1}{4}}=\pm\sqrt{\braket{x^2}_{\text{ss}}},
\label{eq:xpm}
\end{eqnarray}
i.e.~ the equilibria correspond to the mean values found in Eq.~\eqref{eq:x2_wigner}. To corroborate, we calculate the
mean values implied by Eq.~\eqref{eq:Wpeaks},
\begin{eqnarray}
\braket{x^2}_{W\text{ss}}=\sqrt{\frac{A_t+A_p+D_p}{72\Gamma_f}}\frac{\Gamma\left[\frac{5}{4}\right]
\Gamma\left[\frac{\gamma_f}{12\Gamma_f}-\frac{3}{4}\right]}{\Gamma\left[\frac{3}{4}\right]
\Gamma\left[\frac{\gamma_f}{12\Gamma_f}-\frac{1}{4}\right]},
\end{eqnarray}
where $\Gamma[m]$ is the Gamma function with a variable $m$. At optimum feedback \cite{Rodenburg:16},
$\gamma_f=18\Gamma_f$, we find $\braket{x^2}_{W\text{ss}}=\sqrt{3}\sqrt{\frac{A_t+A_p+D_p}{2J}}$ which is
$\sqrt{3}$ times larger than the quantity calculated in Eq.~\eqref{eq:x2_wigner} using the full FP equation. The explanation
for this difference derives from the assumption of overdamped motion, which significantly reduces the probability
of finding the nanoparticle around $x=0$. With this assumption therefore, the moment of $\braket{x^2}_{W\text{ss}}$
is greater than the quantity calculated without the overdamped approximation. The analytical result of
$x_{\pm}$ [Eq.~\eqref{eq:xpm}] agrees quite well with the locations of the peaks of the position distribution
as can be seen from the dashed vertical lines in Fig.~\ref{fig:Wx}.

\begin{figure}[tbp]
\centering
\includegraphics[width=0.9\linewidth]{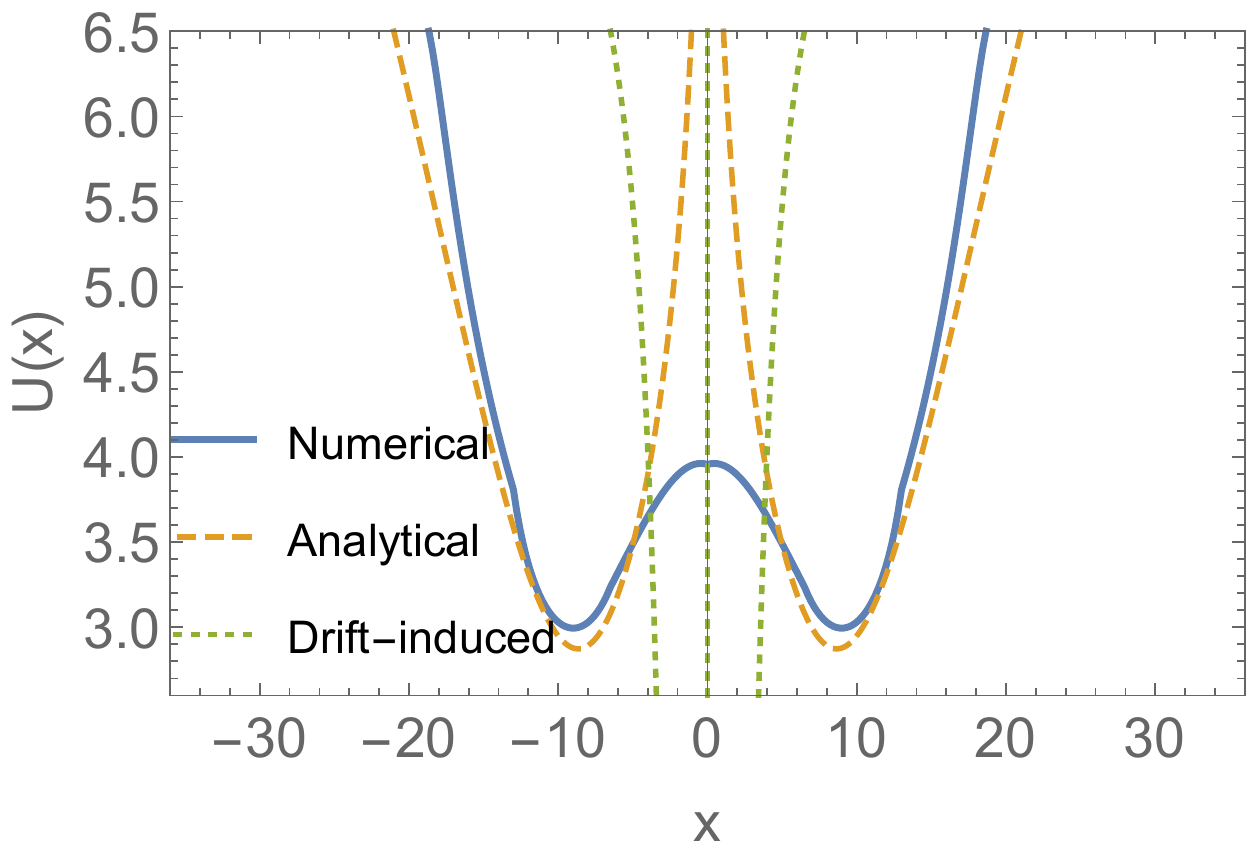}
\caption{The non-equilibrium nanoparticle potential $U(x)$ [Eq.~\eqref{eq:Ux}] as a function of scaled
nanoparticle displacement $x/x_{0}$. The solid curve is the result of numerically solving Eq.~\eqref{eq:Wigner_tr}, 
the dashed curve corresponds to the analytical result found using Eq. \eqref{eq:Wignerx}, and the dotted curve is 
drift-induced potential obtained using $h(x,t)$ from Eq.~\eqref{eq:sm} for the same parameters in Fig.~\ref{fig:Wigner_64} }
\label{fig:BistablePot}
\end{figure}

In order to understand the bistability from the perspective of energy, we consider the non-equilibrium nanoparticle
potential \cite{Kus:83}
\begin{equation}
U(x)=-k_{B}T\mathrm{ln}[W_{x\text{ss}}(x)],
\label{eq:Ux}
\end{equation}
where $\mathrm{ln}[x]$ is the natural logarithm of x. We show the analytical potential found by combining Eq.~\eqref{eq:Ux} and Eq. \eqref{eq:Wignerx} as well as the
corresponding numerical calculation from Eq.~\eqref{eq:Wigner_tr} in Fig.~\ref{fig:BistablePot}. Both curves show that the potential $U(x)$ exhibits double minima that correspond to nanoparticle bistability. The analytical curve overestimates the barrier between the two wells due to the assumption of overdamping. The
numerical curve presents a lower barrier.

We emphasize that in contrast to cavity optomechanics \cite{Dorsel:83,Meystre:85,Metzger:08,Ghobadi:11,Aspelmeyer:14},
bistability in our model cannot be explained by solely considering the mean effects of the feedback and neglecting 
the fluctuating terms in the Langevin equation. This can be seen by plotting the potential of Eq.~\eqref{eq:Ux} 
obtained by setting $g(x,t)=0.$ This potential, which is entirely due to the nonfluctuating contribution $h(x,t)$ 
in the Langevin equation of Eq.~\eqref{eq:sm}, is shown as a dotted curve in Fig.~\ref{fig:BistablePot} and clearly 
does not exhibit the shape required to explain bistability. For the parameters used to calculate the full potential which includes the stochastic processes of the system, we find the gas fluctuations
represented by $D_{p}$ form the dominant diffusive contribution to the bistability, while the effects of optical
scattering and feedback backaction are small.

\begin{figure}[tbp]
\centering
\includegraphics[width=0.9\linewidth]{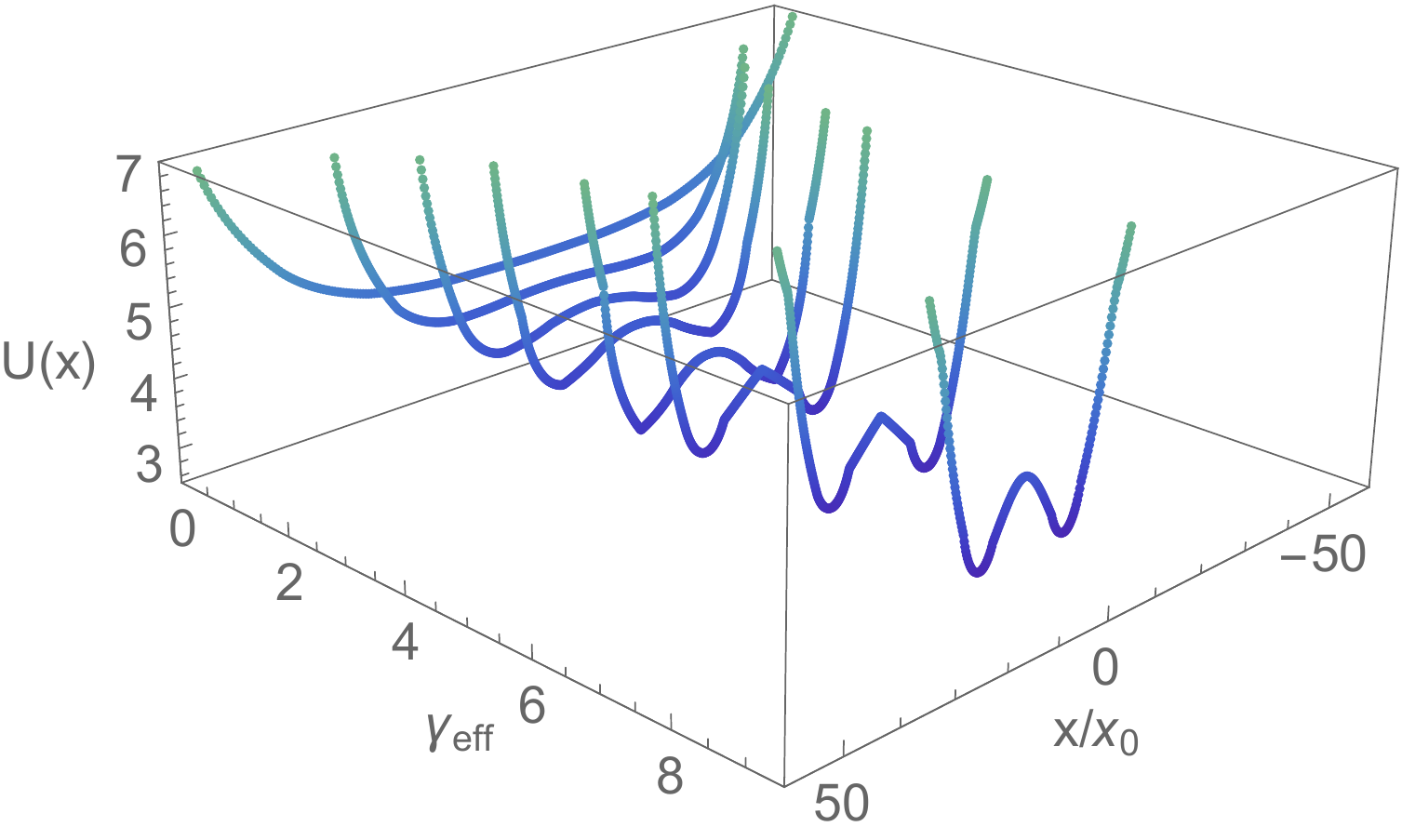}
\caption{Full non-equilibrium nanoparticle potential as a function of feedback strength $\gamma_{\text{eff}}$. As the feedback
increases, the potential changes from simple harmonic to bistable. The other parameters are the same as in Fig.~\ref{fig:Wigner_64}.}
\label{fig:BistablePot2}
\end{figure}

\begin{figure}[tbp]
\centering
\includegraphics[width=0.95\linewidth]{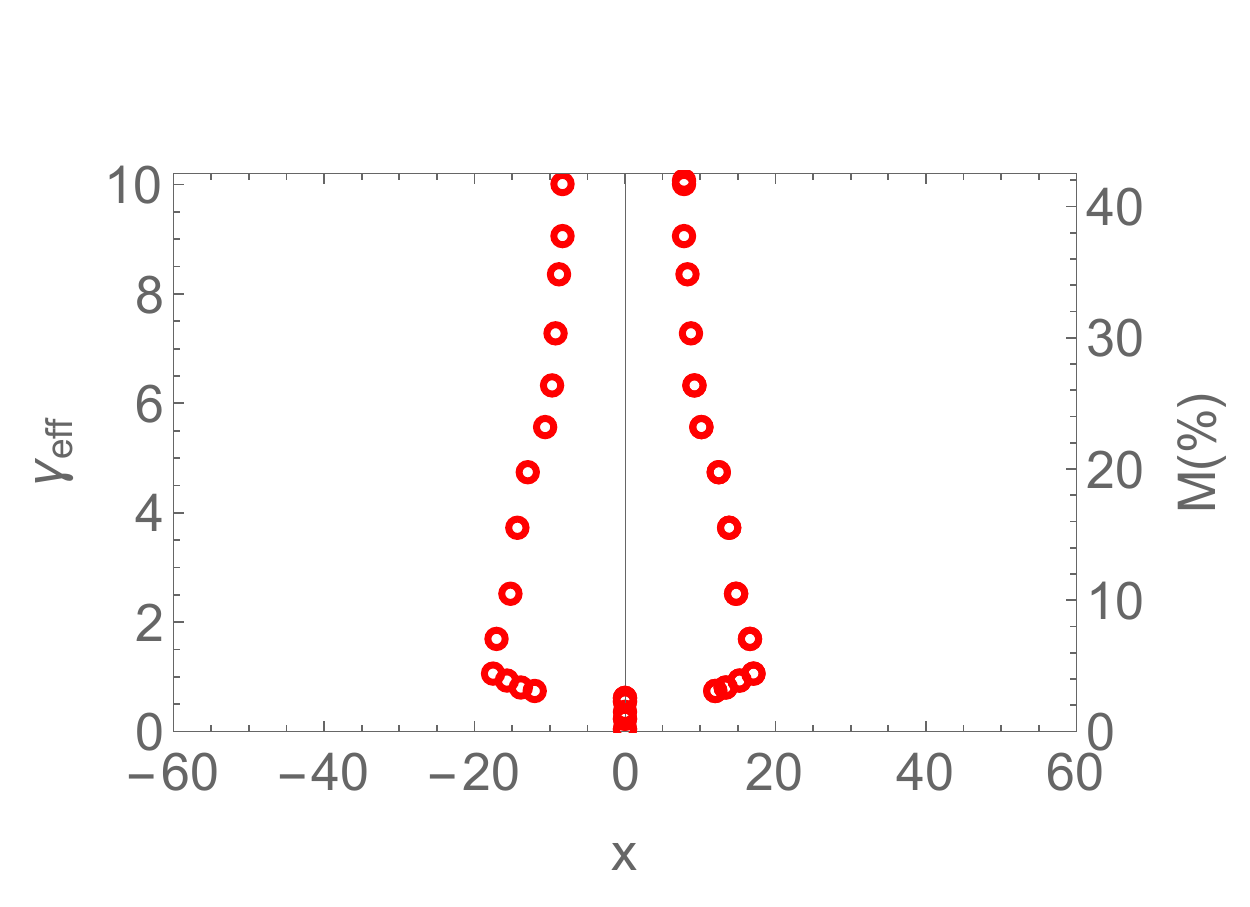}
\caption{The maxima in the nanoparticle position distributions as a function of feedback strength $\gamma_{\text{eff}}$.
As the feedback increases, the monostable solution turns into a bistable solution. The modulation of the trap laser
beam intensity is also indicated on the right vertical axis. The other parameters are the same as in Fig.~\ref{fig:Wigner_64}.}
\label{fig:BistablePot3}
\end{figure}

In order to show how the bistability turns on, we have also plotted in Fig.~\ref{fig:BistablePot2} the
evolution of the non-equilibrium nanoparticle potential $U(x)$, obtained by solving Eq.~\eqref{eq:Wigner_tr}, as a
function of the feedback cooling strength $\gamma_{\text{eff}}$. As can be seen, with increasing feedback, the potential
changes from simple harmonic to bistable.
The corresponding plot for the maxima of the nanoparticle position distribution as a function of $\gamma_{\text{eff}}$ is
shown in Fig.~\ref{fig:BistablePot3}.
It should be noted while considering these plots that the maximum allowed value of the feedback strength is restricted
by the modulation of the trap laser beam intensity \cite{Rodenburg:16},
\begin{equation}
M=\frac{\gamma_{f}\braket{n}_{\text{ss}}}{\omega_{z}},
\end{equation}
which cannot be more than $100\%$. Also we note here that we presented feedback-induced bistability for low steady-state phonon numbers ($\braket{n}_{\text{ss}}<100$). However, bistability may be obtained even for high phonon numbers as long as the condition Eq.~\eqref{eq:overdampc} is satisfied. For example, we have verified the presence of bistability for $\gamma_g=0.5$ kHz, $\gamma_f=3.3\times10^{-5}$ kHz, $\Gamma_f=2.8\times10^{-10}$ kHz, which imply $\gamma_{\text{eff}}=9.9$ and $\braket{n}_{\text{ss}}=3.53\times10^5$, conditions that may be easier to realize experimentally. 

We conclude this section by stating one advantage of the FP approach, which is that any moment can be calculated
directly by performing integration using the Wigner function. This is important especially for finding higher-order
moments which are difficult to obtain directly using the master equation Eq.~\eqref{eq:master_com}, as pointed out
in Section \ref{subsec:WPF}. For example, for optimum feedback, we obtain the fourth order steady-state moment by
integration
\begin{eqnarray}
\braket{x^4}_{W\text{ss}}=\frac{9(A_t+A_p+D_p)}{2J}=3\braket{x^2}_{W\text{ss}}^2,
\end{eqnarray}
which is the same as that for a zero-mean Gaussian distribution and was assumed for the position distribution in
Section \ref{subsec:WPF}.

\subsubsection{The momentum distribution}
\begin{figure}[tbp]
\centering
\includegraphics[width=0.95\columnwidth]{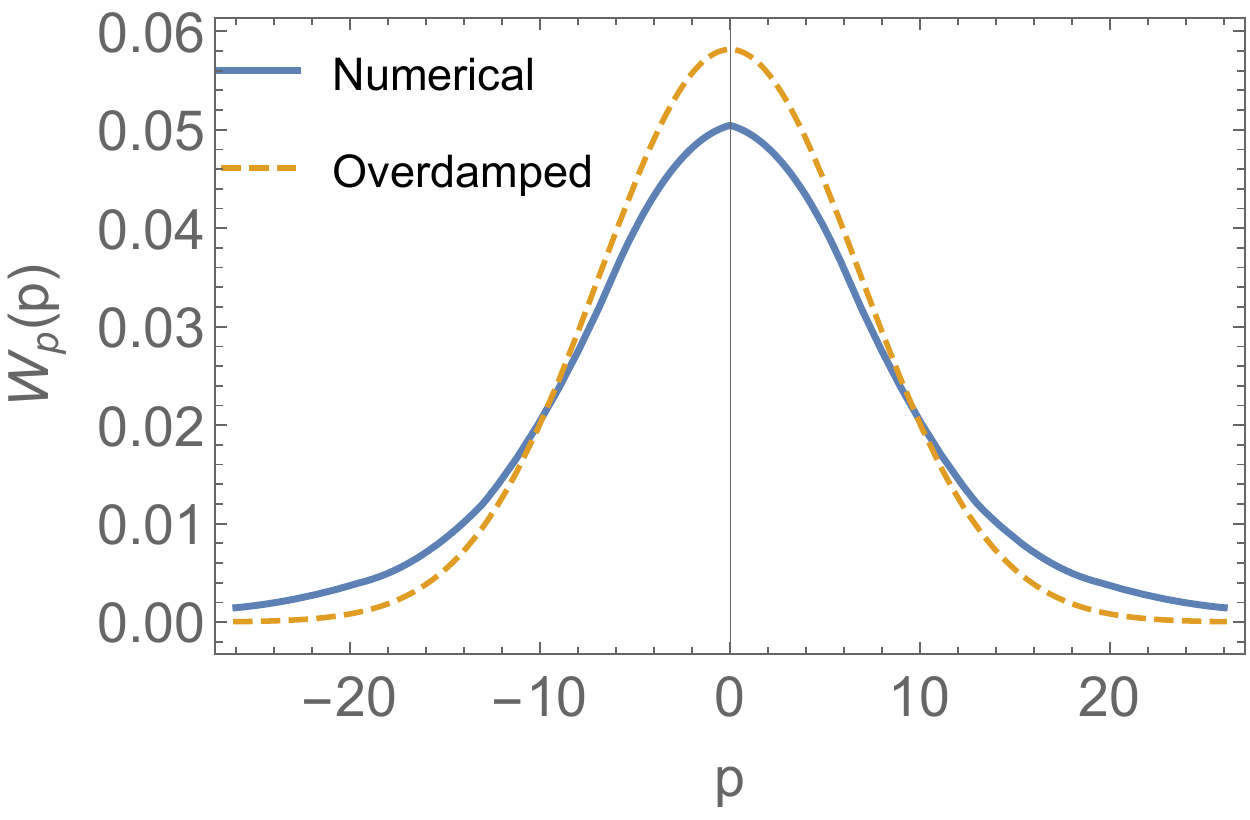}
\caption{Steady-state momentum distributions from the numerical solution (solid) and that using $W_{p\text{ss}}(p)$
(dashed) for the same parameters in Fig.~\ref{fig:Wigner_64}. \label{fig:Wp}}
\end{figure}

We now consider the momentum distribution in the overdamped regime. Interestingly, it turns out to be analytically
more accessible than the position distribution, thus making the calculation of higher moments more amenable. By
performing an integration on $x$ on both sides of Eq.~\eqref{eq:Wigner_tr}, we obtain the following FP equation
for the momentum distribution
\begin{eqnarray}
\frac{\partial W_p(p,t)}{\partial t}=-\frac{\partial }{\partial p}\left(D_p^{(1)}W_p(p,t)\right)+\frac{\partial^2 }{\partial p^2}\left(D_p^{(2)}W_p(p,t)\right),\nonumber\\
\end{eqnarray}
where $D_p^{(1)}=-2\gamma_gp-24\gamma_f\braket{x^2}_{W\text{ss}}p$ and
$D_p^{(2)}=(A_t+A_p+D_p)+72\Gamma_f \braket{x^4}_{W\text{ss}}$.
In the steady-state, the momentum distribution can be
obtained analytically in the form of a Gaussian function given by
\begin{equation}
W_{p\text{ss}}(p)=\mathcal{N}_{p}\exp{\left(-\frac{\gamma_g+12\gamma_f\braket{x^2}_{W\text{ss}}}{(A_t+A_p+D_p)+72\Gamma_f \braket{x^4}_{W\text{ss}}}p^2\right)},
\label{eq:Wp}
\end{equation}
where $\mathcal{N}_{p}$ is a normalization constant. The momentum distribution of Eq.~\eqref{eq:Wp} as well as the
corresponding quantity calculated from Eq.~\eqref{eq:Wigner_tr} are shown in Fig.~\ref{fig:Wp}, and agree quite well
with each other. We obtain from Eq.~\eqref{eq:Wp} the second and fourth order moments as
\begin{eqnarray}
\braket{p^2}_{W\text{ss}}&=&\frac{(A_t+A_p+D_p)+72\Gamma_f \braket{x^4}_{W\text{ss}}}{2\gamma_g
+24\gamma_f\braket{x^2}_{W\text{ss}}}\nonumber\\ &=&\sqrt{\frac{A_t+A_p+D_p}{3J/2}},\\
\braket{p^4}_{W\text{ss}}&=&3\braket{p^2}_{W\text{ss}}^2=2\left(\frac{A_t+A_p+D_p}{J}\right).
\end{eqnarray}
From these expressions, we obtain the steady-state mean phonon number
\begin{equation}
\braket{n}_{W\text{ss}}=\frac{5}{2\sqrt{3}}\sqrt{\frac{A_t+A_p+D_p}{2J}}-\frac{1}{2},
\label{eq:nWss}
\end{equation}
which qualitatively agrees with the result of Eq.~\eqref{eq:x2_wigner}. For the parameters of Fig.~\ref{fig:Wigner_64},
we calculate the mean phonon number $\braket{n}_{\text{ns}}=84$ using the numerical solution to Eq.~\eqref{eq:Wigner_tr}.
This result is close to the analytical results $\braket{n}_{\text{ss}}=65$ [Eq.~\eqref{eq:steady_n}] and
$\braket{n}_{\text{Wss}}=75$ [Eq.~\eqref{eq:nWss}]. The mean of the phonon number squared is given by
\begin{eqnarray}
\braket{n^2}_{W\text{ss}}&=&\int W_{\text{ss}}(x,p)\left[\left( \frac{x^2+p^2}{2}\right)^2
-\left(\frac{x^2+p^2}{2}\right)\right]dx dp\nonumber\\
&\approx&\frac{1}{4}\left(\braket{x^4}_{W\text{ss}}+2\braket{x^2}_{W\text{ss}}\braket{p^2}_{W\text{ss}}
+\braket{p^4}_{W\text{ss}}\right)\nonumber\\
&&-\frac{1}{2}\left(\braket{x^2}_{W\text{ss}}+\braket{p^2}_{W\text{ss}}\right)\nonumber\\
&=&\frac{17}{8}\frac{A_t+A_p+D_p}{J}-\frac{5}{2\sqrt{3}}\sqrt{\frac{A_t+A_p+D_p}{2J}}.\nonumber\\
\label{eq:mean_n2}
\end{eqnarray}
Therefore, the second-order correlation function for the feedback-cooled nanoparticle in the overdamped regime
is given by
\begin{eqnarray}
g^{(2)}&=&\frac{\braket{n^2}_{W\text{ss}}-\braket{n}_{W\text{ss}}}{\braket{n}_{W\text{ss}}^2}\nonumber\\
&\approx&\frac{51}{25},
\end{eqnarray}
where we consider the overdamped condition $\braket{x^2}_{W\text{ss}}\gg1$ to obtain the last line. This result
shows that in the overdamped regime the nanoparticle has a phonon bunching relation close to a thermal
mechanical state for which $g^{(2)}=2$ \cite{Rodenburg:16}.

\section{Conclusion}
\label{sec:Conc}
In conclusion, we have studied the low and high damping regimes for an optically levitated nanoparticle subject to
parametric feedback. We have used a Fokker Planck treatment, useful for viewing the phase-space distribution of the
nanoparticle and calculating higher-order moments of the oscillator variables in the presence of nonlinear feedback.
For low damping, we obtained the position and energy distributions of Ref. \cite{Gieseler:14nn, Gieseler:15}, but
starting from a fully quantum mechanical model. For high damping, we found a feedback-induced bistability.
It is important to note that unlike in the case of cavity-based systems, the bistability manifests itself only in the
mechanical, and not optical, degree of freedom. Also, the bistability is crucially influenced by system fluctuations 
and therefore cannot be described using a classical mean value equation as in standard cavity optomechanics. Our 
prediction should be experimentally observable as we have used realistic parameters in our analysis. Using our 
formalism we have also investigated energy equipartition, phonon number behavior, and phonon bunching for the 
nanoparticle, which shows close resemblance to that of a thermal mechanical state. We presented both analytical and 
numerical results to support our conclusions. Our study opens the door to the use of bistability in cavity-free 
levitated systems, and provides a framework for studying the phase-space properties of such systems.

\section*{ Acknowledgments}
This research is supported by the Office of Naval Research under Award No. N00014-14-1-0803. We thank
N. Vamivakas, R. Pettit, and L. Neukirch for useful discussions.




\end{document}